\newcommand{\home}{n:}          

\documentclass[a4paper,12pt,titlepage]{article}

\usepackage{epsfig}

\usepackage{acronym}
\acrodef{MRC}{Medical Research Council}
\acrodef{BSU}{\ac{MRC} Biostatistics Unit}
\usepackage{amsmath,amsthm}
\usepackage{breqn}
\usepackage{natbib}
\usepackage{hyperref}
\usepackage{xcolor}
\usepackage{threeparttable}
\usepackage{multirow}
\usepackage{graphicx}
\usepackage{titling}

\newcommand {\E}      [1] {{\mbox{E}}\left[#1\right]}
\newcommand {\varb}   [1] {{\rm var}\left(#1\right)}

\newcommand {\logit}  [1] {{\rm logit}\ #1}
\newtheorem{assumption}{Assumption}

\newcommand {\pret}{\le{t}}
\newcommand {\postt}{>t}

\newcommand{\bY}{\boldsymbol{Y}}
\newcommand {\tmax}{t_{max}}

\renewcommand{\baselinestretch}{1.5}

\title{A causal modelling framework for reference-based imputation and tipping point analysis}
\author{Ian R. White $^{1,2}$ \and Royes Joseph $^1$ \and Nicky Best $^3$ \\
$^1$ MRC Biostatistics Unit, University of Cambridge, UK \\
$^2$ MRC Clinical Trials Unit at UCL, London, UK \\
$^3$ GlaxoSmithKline, Uxbridge, UK
}
\date{\today\\[5ex]}

\begin{document}

\renewcommand{\baselinestretch}{1.5}


\maketitle



\begin{abstract} 
We consider estimating the ``de facto'' or effectiveness estimand in a randomised placebo-controlled or standard-of-care-controlled drug trial with quantitative outcome, where participants who discontinue an investigational treatment are not followed up thereafter.
\citet{Carpenter++13} proposed reference-based imputation methods which use a reference arm to inform the distribution of post-discontinuation outcomes and hence to inform an imputation model. However, the reference-based imputation methods were not formally justified. We present a causal model which makes an explicit assumption in a potential outcomes framework about the maintained causal effect of treatment after discontinuation. We show that the ``jump to reference'', ``copy reference'' and ``copy increments in reference'' reference-based imputation methods, with the control arm as the reference arm, are special cases of the causal model with specific assumptions about the causal treatment effect. Results from simulation studies are presented. We also show that the causal model provides a flexible and transparent framework for a tipping point sensitivity analysis in which we vary the assumptions made about the causal effect of discontinued treatment. We illustrate the approach with data from two longitudinal clinical trials.


\end{abstract}




\clearpage

\section{Introduction}

Missing outcome data represent a major threat to the validity of randomised clinical trials (RCTs), and appropriate analysis methods have been much discussed. An influential report showed that different analysis methods may target different estimands (different measures of treatment effect) and argued that specification of the estimand is an important part of trial design and should inform trial analysis and reporting \citep{CNSTAT10}. Regulators have joined the call for estimands to be defined clearly \citep{ICHE9}.

Estimands have long been discussed in the causal inference literature. For example, \citet{Angrist++96} discussed all-or-nothing non-compliance in RCTs and gave conditions under which instrumental variable methods estimate the complier-average causal effect (CACE), that is, the mean effect of treatment in those who will receive treatment if and only if they are assigned to treatment. This differs from the intention-to-treat estimand which is the mean effect of allocation to treatment. In general, the estimand can be designed to take into account or not take into account potential confounding of the data by post-randomisation events such as non-compliance, discontinuation of intervention, treatment switching, or use of rescue medication.

We consider two types of estimand considered by \citet{CNSTAT10}: (E1) difference in outcome improvement at the planned endpoint if all participants had tolerated or adhered to trial protocol; (E2) difference in outcome improvement at the planned endpoint for all randomised participants. The former measures how treatment works in an ideal setting (efficacy),
while the latter measures how treatment might work in practice (effectiveness).
To encompass outcomes that measure harms of treatment, \citet{Carpenter++13} proposed the broader terms \textit{de jure} and \textit{de facto} estimand for (E1) and (E2) respectively.

Sometimes investigators continue to collect data after treatment discontinuation, which we call ``off-treatment data". The use of such data depends on the estimand \citep{Permutt2016a}.
For the estimation of a \textit{de jure} estimand, off-treatment data for participants who discontinued randomised treatment could be used in a CACE analysis \citep{Dunn++03}. In practice, however, off-treatment data are typically either not collected or are excluded from the primary analysis, and the missing data are assumed to be missing at random (MAR): that is, it is assumed that participants who discontinued treatment would have benefited from continued treatment in the same way as those who remained on treatment. However, estimation of a \textit{de facto} estimand ideally makes use of the off-treatment data, and guidance recommends its collection where possible \citep{CNSTAT10}. When all discontinuers are followed up and outcome data obtained, the \textit{de facto} estimand can be directly estimated by comparing observed means \citep{LittleKang15}.

This paper considers estimation of a \textit{de facto} estimand for a quantitative outcome in the absence of off-treatment data. This requires assumptions about how participants who have discontinued treatment benefit from their previous treatment. For example, participants who received only a partial period of treatment may not have benefited after treatment discontinuation in a similar manner to those who completed the treatment in the trial. We also focus on the situation in which rescue treatment (over and above the per protocol treatment regime for the control arm) for those who discontinue randomised treatment is not available, or interest is in the effect attributable to the initially randomised treatment without the confounding effects of rescue medications (where the latter corresponds to estimand 6 in \citet{mallinckrodt2012}).

In previous work on this topic, \citet{LittleYau96} presented a multiple imputation (MI) approach that can incorporate a variety of alternative assumptions about treatment dose after treatment discontinuation for the estimation of \textit{de facto} estimands in RCTs.
\citet{Carpenter++13} presented a generic algorithm for MI of post-discontinuation outcome data. They derived the conditional distribution of post-discontinuation outcome data given pre-discontinuation outcome data and time of discontinuation from a joint Normal distribution which could be formed with reference to other arms in several ways.
For the ``copy reference'' (CR) method, the mean of the joint distribution is taken as the mean for the reference arm.
For other methods, the mean of the joint distribution in a specific arm follows the mean of that arm until treatment discontinuation;
after treatment discontinuation, it is formed either
by directly using the means over time for the reference arm (the ``jump to reference'' (J2R) method);
by using the mean in the specific arm at the time of treatment discontinuation (the ``last mean carried forward'' (LMCF) method);
or
by using the mean for the specific arm at the time of treatment discontinuation plus the mean increments over time in the reference arm (the ``copy increments in reference'' (CIR) method).
The variance-covariance matrix of the joint distribution is set equal to the variance-covariance matrix from either the active (investigational) treatment arm or the reference arm. However, \citet{Carpenter++13} did not provide a theoretical justification of these ``reference-based imputation'' (RBI) methods.

Specification of estimands is clarified by using counterfactual outcomes -- outcomes that have not or could not have been observed. Such counterfactuals are best described using potential outcomes notation \citep{LittleRubin00,Angrist++96}. Our aims in this paper are to provide mathematical derivation of the RBI methods, including their identifying assumptions, in a potential outcomes framework, and to propose a post-discontinuation imputation model using explicit causal assumptions to enable generalization of the RBI methods.
We do this by assuming that after discontinuing their randomised treatment, participants receive treatment that is similar to that allocated to the control arm: thus in the case of reference-based imputation, we take the reference to be the control treatment, where this is typically either placebo or standard of care (SoC). Further, while we allow both a delayed response to the active treatment and a maintained response after discontinuing the active treatment, we implicitly assume that there is no delayed response to the control treatment: thus when a patient discontinues randomised treatment, we assume the effects of any treatments they switch to are similar to the effects they would have experienced had they received the control treatment from the start of the trial.

Since the imputation models in our proposal rely on untestable assumptions, sensitivity analyses are desirable for understanding the impact of the assumptions on inferences and conclusions from the primary analysis. \citet{Kenward++01} described a principled approach to sensitivity analyses  which varies a sensitivity parameter that quantifies deviations from the missing data assumption. A tipping point sensitivity analysis (e.g. \citet{Yan2009,LiublinskaRubin14}) extends this approach by varying the sensitivity parameter until the conclusion from the primary analysis is overturned. In this paper, we propose a tipping point sensitivity analysis using the causal model.

In Section 2, we investigate the underlying assumptions for RBI methods and present a causal model that requires an explicit assumption on post-discontinuation behaviour in terms of the average treatment effect which is maintained after treatment discontinuation. In Section 3, we demonstrate the equivalence of RBI and causal model estimates in a simulation study. In Section 4, we illustrate the proposed approach and demonstrate the tipping point analysis with two example data sets. We conclude with summary remarks in Section 5.

\section{RBI methods and causal model}\label{sec:methods}

Our aim here is to propose a causal model which can be used to impute missing values, and to relate the mean imputations from the causal model to the mean imputations derived under RBI. We start by presenting notation and estimands, and then present the assumptions used in imputation, a comparison of the methods, and details of implementation.

\subsection{Notation and simplifying assumptions}

We consider a two-arm longitudinal RCT with $\tmax$ scheduled observation times after randomisation.
Let $Z$ denote the randomised treatment arm with $Z=a$ for the active treatment and $Z=c$ for the control treatment.
Let $Y_t$ denote the outcome at the $t$th observation time, where $t =0,1,...,\tmax$ and $t=0$ denotes the baseline observation.
Each $Y_t$ may be observed or missing; if missing, it is the value of the actual outcome that is experienced by the participant despite being unobserved by the investigators.
Let $\boldsymbol{Y} = (Y_0,...,Y_\tmax)^'$ denote the vector of all outcomes, with subvectors
$\boldsymbol{Y}_{\pret} = (Y_0,...,Y_t)^'$ for outcomes until time $t$ (including the baseline), and
$\boldsymbol{Y}_{\postt} = (Y_{t+1},...,Y_\tmax)^'$ for outcomes after time $t$.
We assume for clarity that the only baseline covariate is the baseline value of the outcome, $Y_0$, but other baseline covariates can be accommodated by replacing $Y_0$ with a baseline covariate vector $\bY_0$.

We assume that individuals in both arms may discontinue their randomised treatment during the course of follow-up, but they do not restart treatment. We let $D_a$, $0 \le D_a \le \tmax$, denote the last observation time for an active arm participant prior to stopping treatment, and similarly for $D_c$ in the control arm. $D_a$ or $D_c=\tmax$ denotes a participant who continues treatment throughout the trial.
We assume that, by the design of the trial, there is no follow-up after treatment discontinuation, and that data may also be intermittently missing before treatment discontinuation.

We define the potential outcome $Y_t(s)$, for $0 \le s,t \le \tmax$, as the true outcome at time $t$ if the active treatment were received for $s$ periods and then control treatment were received.
We assume that outcomes are not causally affected by future treatment, so
\begin{equation}\label{eq:Ydefn}
Y_t(s)=Y_t(t) \mbox{ for all } s>t.
\end{equation}
In particular, (\ref{eq:Ydefn}) means that the baseline value is unaffected by treatment, $Y_0(s)=Y_0$ for all $s$.
All counterfactual outcomes $Y_t(s)$ are assumed to be independent of $Z$ (by randomisation), but not of $D_a$ or $D_c$ (treatment discontinuation may be related to counterfactual outcomes).
Let $\boldsymbol{Y}(s) = (Y_{0}(s),...,Y_\tmax(s))^'$ with subvectors
$\boldsymbol{Y}_{\pret}(s) = (Y_{0}(s),...,Y_t(s))^'$
and $\boldsymbol{Y}_{\postt}(s) = (Y_{t+1}(s),...,Y_\tmax(s))^'$.
The overall means of counterfactuals are defined as $\boldsymbol{\mu}(s)=\E{\boldsymbol{Y}(s)}$, $\boldsymbol{\mu}_{\pret}(s)=\E{\boldsymbol{Y}_{\pret}(s)}$ and $\boldsymbol{\mu}_{\postt}(s)=\E{\boldsymbol{Y}_{\postt}(s)}$.

Finally, let the variance-covariance matrices be
$\boldsymbol{\Sigma}(s) = \varb{\boldsymbol{Y}(s)}$ for $s=0$ or $\tmax$, with submatrices $\boldsymbol{\Sigma}_{\pret\pret}(s)$, $\boldsymbol{\Sigma}_{\postt\pret}(s)$, $\boldsymbol{\Sigma}_{\pret\postt}(s)$, $\boldsymbol{\Sigma}_{\postt\postt}(s)$.
For the linear regression of $\boldsymbol{Y}_{\postt}(s)$ on $\boldsymbol{Y}_{\pret}(s)$, we write the  regression coefficients as
$\boldsymbol{\beta}_t(s)=\boldsymbol{\Sigma}_{\postt\pret}(s) \boldsymbol{\Sigma}_{\pret\pret}(s)^{-1}$
and the residual variance as
$\boldsymbol{\Sigma}_{\postt\postt|\pret}(s) = \boldsymbol{\Sigma}_{\postt\postt}(s) - \boldsymbol{\Sigma}_{\postt\pret}(s) \boldsymbol{\Sigma}_{\pret\pret}(s)^{-1} \boldsymbol{\Sigma}_{\pret\postt}(s)$.
The notation is summarised in Table \ref{tab:notation}.

\begin{center}
[TABLE \ref{tab:notation} ABOUT HERE.]
\end{center}


\subsection{Estimands}
Given this notation, we can write the \textit{de jure} estimand (estimand E1) at time $t>0$ as $E[Y_t(t)-Y_t(0)]$.
The \textit{de facto} estimand (estimand E2) is the estimand of interest in this paper and is defined as $E[{Y}_t(D_a)-{Y}_t(0)]$.
Often, primary interest is in the last observation time, $t=\tmax$.

\subsection{Imputation}

Our aim is to impute $\bY_\postt(t)$ in the $D_a=t$ subgroup where $t<\tmax$.
Because our main concern is bias, we focus on the mean of the imputation distribution. Variances are considered in Section \ref{ssec:impl}.

\subsubsection{Initial assumptions and modelling}


\begin{assumption}\label{ass:refMAR}
  $(\boldsymbol{Y} | Z=c)$ is missing at random (MAR).
\end{assumption}
Assumption  \ref{ass:refMAR} states that the observed outcomes in the control arm are MAR.
If there are no missing data before treatment discontinuation then we can also write this
  $$p(D_c=t|Z=c, \bY, D_c \ge t) = p(D_c=t|Z=c, \bY_{\pret}, D_c \ge t)$$
for all $t$. In other words, treatment discontinuation in the control arm does not depend on future (untreated) outcomes, given the past and present.

\begin{assumption}\label{ass:actMAR1}
  $(\boldsymbol{Y}(\tmax) | Z=a)$ is MAR.
\end{assumption}
Assumption \ref{ass:actMAR1} states that the counterfactual fully treated outcomes up to time $t$ in the active arm are MAR.
If there are no missing data before treatment discontinuation then we can also write this
  $$p(D_a=t| Z=a, \bY(\tmax),D_a \ge t)
  = p(D_a=t|Z=a, \bY_{\pret}(\tmax), D_a \ge t)$$
for all $t$.
In other words, treatment discontinuation in the active arm does not relate to future fully-treated outcomes, given the past and present.

\textbf{We do not assume} that the actual outcomes in the active arm, $(\boldsymbol{Y}| Z=a)$, are MAR.
Indeed, this is unlikely to be true, since (if treatment works) treatment discontinuation causally affects actual future outcomes. Thus treatment discontinuation is allowed to relate to future \emph{actual} outcomes, given the past and present.

Assumption \ref{ass:refMAR} allows us to estimate $\boldsymbol{\mu}(0)$ and $\boldsymbol{\Sigma}(0)$ by applying likelihood-based methods to the observed outcomes in arm $Z=c$.
Assumption \ref{ass:actMAR1} allows us to estimate $\boldsymbol{\mu}(\tmax)$ and $\boldsymbol{\Sigma}(\tmax)$ by applying likelihood-based methods to the observed outcomes in arm $Z=a$.
These two assumptions are therefore enough to identify the \textit{de jure} estimand.
They also allow us to impute intermittent missing data before treatment discontinuation in the active treatment arm, and all missing data in the control arm, so henceforth we assume that the only missing data are actual outcomes after treatment discontinuation in the active arm.

\subsubsection{Post-discontinuation imputation using explicit causal assumptions}

For the \textit{de facto} estimand, we need to impute actual outcomes after treatment discontinuation.
We consider an individual in the active arm who stops treatment at time $t<\tmax$ with history $\boldsymbol{Y}_{\pret}$.
We use as imputation model a multivariate linear regression model with variance-covariance matrix described in Section \ref{ssec:impl}. We write the mean in this model as the sum of three terms:
\begin{eqnarray}\label{eq:mumisparts}
E[\boldsymbol{Y}_{\postt}(t) | Z=a, \boldsymbol{Y}_{\pret}, D_a=t]
&=& \left\{ E[\boldsymbol{Y}_{\postt}(t) | Z=a, \boldsymbol{Y}_{\pret}, D_a=t] - \boldsymbol{\mu}_{\postt}(t) \right\}\nonumber\\
&&+ \left\{\boldsymbol{\mu}_{\postt}(t) - \boldsymbol{\mu}_{\postt}(0)\right\}
+ \boldsymbol{\mu}_{\postt}(0)
\end{eqnarray}
where the first term represents the difference between the subgroup who discontinue at time $t$ and the whole group (``selection term''), the second term represents the treatment effect in the whole group (``maintained treatment effect''), and the third term is the mean in the control arm. The third term is identified by assumption \ref{ass:refMAR}, and we make the identifying assumptions below for the first two terms on the right hand side of (\ref{eq:mumisparts}).

\subsubsection{Selection term}

To identify the first term in (\ref{eq:mumisparts}), we assume
\begin{assumption}\label{ass:linear}
   $\boldsymbol{Y}_{\postt}(t) | \boldsymbol{Y}_{\pret}(t)$ follows a linear regression for each $t$.
\end{assumption}

\begin{assumption}\label{ass:actMAR2}
   $p(D_a=t|Z=a, \boldsymbol{Y}(t)) = p(D_a=t|Z=a, \boldsymbol{Y}_{\pret}(t))$.
\end{assumption}

\ref{ass:actMAR2} appears similar to \ref{ass:actMAR1}, but whereas \ref{ass:actMAR1} states that treatment discontinuation at time $t$ is unaffected by future fully-treated potential outcomes, \ref{ass:actMAR2} refers instead to future partly-treated potential outcomes.
If there are no missing data before treatment discontinuation then a stronger assumption which implies both \ref{ass:actMAR1} and \ref{ass:actMAR2} is
$$
p(D_a=t | Z=a, \boldsymbol{Y}(s), D_a \ge t)
=
p(D_a=t | Z=a, \boldsymbol{Y}_{\pret}(s), D_a \ge t)
$$
for all $t$ and all $s>t$.

We can now write the selection term as
\begin{equation}\label{eq:sel}
\begin{array}[b]{llll}
\E{\boldsymbol{Y}_{\postt}(t) | Z=a, \boldsymbol{Y}_{\pret}(t), D_a=t} &-& \boldsymbol{\mu}_{\postt}(t) &  \\
  \hspace{2 em}= \E{\boldsymbol{Y}_{\postt}(t) | Z=a, \boldsymbol{Y}_{\pret}(t)} &-& \boldsymbol{\mu}_{\postt}(t) & \mbox{ (by \ref{ass:actMAR2})} \\
  \hspace{2 em}= \E{\boldsymbol{Y}_{\postt}(t) | \boldsymbol{Y}_{\pret}(t)} &-& \boldsymbol{\mu}_{\postt}(t) & \mbox{ (by randomisation)} \\
  \hspace{2 em}=  \boldsymbol{\beta}_{t}(t) \left\{\boldsymbol{Y}_{\pret}(t) - \boldsymbol{\mu}_{\pret}(t)\right\} &&& \mbox{ (by \ref{ass:linear})}
\end{array}
\end{equation}
where $\boldsymbol{\beta}_{t}(t)$ is the matrix of regression coefficients of $\boldsymbol{Y}_{\postt}(t)$ on $\boldsymbol{Y}_{\pret}(t)$ and is not yet identified.
However $\boldsymbol{\beta}_{t}(0)$, the regression of $\boldsymbol{Y}_{\postt}(0)$ on $\boldsymbol{Y}_{\pret}(0)$,
and $\boldsymbol{\beta}_{t}(\tmax)$, the regression of $\boldsymbol{Y}_{\postt}(\tmax)$ on $\boldsymbol{Y}_{\pret}(t)$, are  identified by assumptions \ref{ass:refMAR} and \ref{ass:actMAR1} respectively.
If all treatment effects are homogeneous (i.e.\ if $\boldsymbol{Y}(t)-\boldsymbol{Y}(0)$ does not vary between individuals for any $t$) then $\boldsymbol{\beta}_{t}(t)=\boldsymbol{\beta}_{t}(\tmax)=\boldsymbol{\beta}_{t}(0)$.
Otherwise, they typically differ.
Often, we may be willing to assume that $\boldsymbol{\Sigma}(\tmax)=\boldsymbol{\Sigma}(0)$, in which case $\boldsymbol{\beta}_{t}(\tmax)=\boldsymbol{\beta}_{t}(0)$ and this is clearly a sensible choice for $\boldsymbol{\beta}_{t}(t)$.
More generally we propose assuming either $\boldsymbol{\beta}_{t}(t)=\boldsymbol{\beta}_{t}(\tmax)$ or $\boldsymbol{\beta}_{t}(t)=\boldsymbol{\beta}_{t}(0)$.

\subsubsection{Maintained treatment effect} \label{sssec:mte}

To identify the second term in (\ref{eq:mumisparts}), we make an explicit assumption about the maintained effect of treatment after discontinuation:
\begin{assumption}\label{ass:del}
   $\E{\boldsymbol{Y}_{\postt}(t)-\boldsymbol{Y}_{\postt}(0)} =
\boldsymbol{K}_t \E{\boldsymbol{Y}_{\pret}(t)-\boldsymbol{Y}_{\pret}(0)}$
\end{assumption}
\noindent or equivalently
\begin{equation}\label{eq:del}
\boldsymbol{\mu}_{\postt}(t)-\boldsymbol{\mu}_{\postt}(0) =
\boldsymbol{K}_t \left\{\boldsymbol{\mu}_{\pret}(t)-\boldsymbol{\mu}_{\pret}(0)\right\}
\end{equation}
where $\boldsymbol{K}_t$ is a $(\tmax - t) \times (t+1)$ matrix of sensitivity parameters: it is not identified by the data and must be specified by the user.
$\boldsymbol{\mu}_{\pret}(0)$ and $\boldsymbol{\mu}_{\pret}(t)$ are identified by assumptions \ref{ass:refMAR} and \ref{ass:actMAR1}, respectively.

Equation (\ref{eq:Ydefn}) implies $\mu_0(t)-\mu_0(0)=0$ so the first column of $\boldsymbol{K}_t$ does not affect equation (\ref{eq:del}) and need not be defined.
In most cases the post-discontinuation treatment effect depends only on the treatment effect at the time of discontinuation and not on treatment effects at earlier times, and therefore $\boldsymbol{K}_t$ needs to have non-zero elements only in the final column: our software implementation below relies on this assumption.
For tipping point sensitivity analyses, we consider two causal models with just one parameter in $\boldsymbol{K}_t$:
\begin{eqnarray}
\E{Y_s(t)-Y_s(0)}&=&k_0 \E{Y_t(t)-Y_t(0)}  \label{eq:mte1} \\
\E{Y_s(t)-Y_s(0)}&=&k_1^{v_s-v_t}\E{Y_t(t)-Y_t(0)} \label{eq:mte2}
\end{eqnarray}
for $s>t$, where $v_s,v_t$ are the times (on a suitable scale) of observations $s,t$.
The maintained treatment effect after treatment discontinuation is constant in model (\ref{eq:mte1}) but decays exponentially in model (\ref{eq:mte2}), being multiplied by $k_1$ in every period, where $0\le k_1 \le 1$.
A combined model is
\begin{eqnarray}
\E{Y_s(t)-Y_s(0)}&=&k_0 k_1^{v_s-v_t}\E{Y_t(t)-Y_t(0)}. \label{eq:mte3}
\end{eqnarray}

\subsection{Comparison with reference-based imputation in active arm} \label{sec:methods:comp}

For  imputing $\boldsymbol{Y}_{\postt}(t)$, \citet{Carpenter++13,Carpenter2014} specified four conditional distributions of post-discontinuation outcomes given  pre-discontinuation outcomes under assumptions \ref{ass:refMAR}, \ref{ass:actMAR1} and \ref{ass:linear}. The means of these imputation distributions in potential outcomes notation are given in Table \ref{tab:imputed} for the $D_a=t$ subgroup for any $t < \tmax$.
In Table \ref{tab:imputed}, $\boldsymbol{C}_t$ is a `carry-forward' $(\tmax-t) \times (t+1)$ matrix containing $t$ columns of zeroes and a final column of ones,
so that $\boldsymbol{C}_t\boldsymbol{\mu}_{\pret}(t)$ is a column vector containing $\tmax-t$ copies of $\mu_t(t)$.
The expression for CR would usually be written $\boldsymbol{\beta}_{t}(0)
\left\{\boldsymbol{Y}_{\pret} -  \boldsymbol{\mu}_{\pret}(0)\right\}
+ \boldsymbol{\mu}_{\postt}(0)$ but has been expanded here to demonstrate similarity with the other expressions.
The corresponding expression for the causal model is derived by substituting equations (\ref{eq:sel}) and (\ref{eq:del}) into (\ref{eq:mumisparts}) and is given in the last row of Table \ref{tab:imputed}.

\begin{center}
[TABLE \ref{tab:imputed} ABOUT HERE]
\end{center}

In these expressions, the first term again represents a selection effect describing how an individual with $D_a=t$ differs systematically from the whole group. The second term represents the effect of treatment up to time $t$ (compared with not receiving the active arm treatment at all) on the outcome, except for LMCF.
From Table \ref{tab:imputed}, it is clear that RBI methods other than LMCF correspond to particular choices of
$\boldsymbol{K}_t$ with $\boldsymbol{\beta}_{t}(t)=\boldsymbol{\beta}_{t}(0)$:
      $\boldsymbol{K}_t=\boldsymbol{0}$ for J2R,
      $\boldsymbol{K}_t=\boldsymbol{C}_t$ for CIR,
      and
      $\boldsymbol{K}_t=\boldsymbol{\beta}_{t}(0) $ for CR.
This makes precise the statement of \citet{mallinckrodt2012} that, under CIR, CR and J2R, the $D_a=t$ subgroup have the treatment effect at time $t$ maintained, diminished and eliminated respectively at time $\tmax$.
The LMCF method does not correspond to the causal model.
Table \ref{tab:imputed} assumes that the variance-covariance matrix is taken from the control arm; it could also be taken from the active treatment arm, in which case the same equivalences apply, but with $\boldsymbol{\beta}_{t}(t)=\boldsymbol{\beta}_{t}(\tmax)$.

A useful expression (derived in \ref{app:defacto}) for the \textit{de facto} estimand at time $\tmax$ is
\begin{eqnarray}
&&E[Y_\tmax(D_a)-Y_\tmax(0)]
= \alpha_\tmax \delta_\tmax + \sum_{t<\tmax} \alpha_t \boldsymbol{e}_t' \boldsymbol{K}_t \boldsymbol{\delta}_t \nonumber \\
&&+ \sum_{t<\tmax} \alpha_t \boldsymbol{e}_t' (\boldsymbol{\beta}_{t}(t)-\boldsymbol{\beta}_{t}(\tmax))
\E{\boldsymbol{Y}_{\pret}(t|D_a=t) - \boldsymbol{Y}_{\pret}(t)} \label{eq:dfest}
\end{eqnarray}
where
$\alpha_t=p(D_a=t)$ expresses the discontinuation distribution in the active treatment arm;
$\delta_t = \E{Y_t(t)-Y_t(0)}$ is the \textit{de jure} estimand, with $\delta_0=0$;
$\boldsymbol{\delta}_t =(\delta_0,\delta_1,\ldots,\delta_t)'$;
and $\boldsymbol{e}_t = (0,...,0,1)'$ of length $\tmax-t$.
Similar expressions apply at times $t<\tmax$. 
From (\ref{eq:dfest}), the bias in the \textit{de facto} estimate associated with wrongly assuming $\boldsymbol{\beta}_t(t)$ to equal $\tilde{\boldsymbol{\beta}}_{t}(t)$ is
$\sum_{t<\tmax} \alpha_t \boldsymbol{e}_t' (\tilde{\boldsymbol{\beta}}_{t}(t)-\boldsymbol{\beta}_{t}(t))\E{\boldsymbol{Y}_{\pret}(t|D_a=t) - \boldsymbol{Y}_{\pret}(t)}$, and
the bias associated with wrongly assuming $\boldsymbol{K}_t$ to equal $\tilde{\boldsymbol{K}_t}$ is
$\sum_{t<\tmax}\alpha_t \boldsymbol{e}_t' (\tilde{\boldsymbol{K}_t}-\boldsymbol{K}_t) \boldsymbol{\delta}_t$.

\subsection{Implementation of the causal model} \label{ssec:impl}

\subsubsection{Option 1: directly using the SAS 5 macros} \label{sssec:sasmacro}

The SAS 5 macros were developed by James Roger to perform multiple imputation under the RBI methods and are available at the web page (on \href{url}{www.missingdata.org.uk}) of the DIA working group for missing data.
We modified the \texttt{Part2A} macro to impute under the causal model with
\begin{equation}\label{eq:Kt}
\boldsymbol{K}_t = k \boldsymbol{C}_t
\end{equation}
where $k$ is a scalar that may vary between participants. This enables causal model (\ref{eq:mte1}) to be implemented by setting $k=k_0$, the same for all participants.
When interest is in the outcome at time $\tmax$, causal model (\ref{eq:mte2}) can be implemented by setting $k=k_1^{\tmax-D_{a}}$, which varies across participants with different values of $D_a$.
The modified macro is available on the DIA working group web page
and sample code is provided in \ref{app:code}.
By default the variance-covariance matrices in the two arms, $\boldsymbol{\Sigma}(\tmax)$ and $\boldsymbol{\Sigma}(0)$, are assumed equal, but the user can specify them to be unequal, which is the case we consider.

\subsubsection{Option 2: indirectly using the SAS 5 macros (based on J2R)}

Equation (\ref{eq:dfest}) for the \textit{de facto} estimand at time $\tmax$ can be written as
\begin{eqnarray*}
E[\boldsymbol{Y}_{\tmax}(D_a)-\boldsymbol{Y}_{\tmax}(0)] = \delta_{J2R} +\sum_{t<\tmax}\alpha_{t}\boldsymbol{e}_t' \boldsymbol{K}_t \boldsymbol{\delta}_t
\end{eqnarray*}
where $\delta_{J2R}$ is the J2R estimand.
Thus we can run the J2R method to estimate the first term, fit a mixed model for repeated measures to estimate the $\delta_t$, directly estimate the $\alpha_t$, and assume values for $\boldsymbol{K}_t$.

To approximate the standard error, we propose using the Rubin's rule (RR) estimate of standard error from J2R, as we will see in Section 3 that the standard error does not vary much with $\boldsymbol{K}_t$.

\subsubsection{Option 3: using Mixed Model Repeated Measures (MMRM) estimates or standard Multiple Imputation (MI)-based estimates}

If we are willing to assume either that $\boldsymbol{\beta}_{t}(t)=\boldsymbol{\beta}_{t}(\tmax)$ or that there is no selection effect, then the third term in expression (\ref{eq:dfest}) vanishes and the \textit{de facto} estimand at time $\tmax$ can be written as
\begin{eqnarray*}
E[\boldsymbol{Y}_{\tmax}(D_a)-\boldsymbol{Y}_{\tmax}(0)] &=& \alpha_{\tmax}\delta_{\tmax}+\sum_{t<\tmax}\alpha_{t}\boldsymbol{e}_t' \boldsymbol{K}_t \boldsymbol{\delta}_t.
\end{eqnarray*}
That is, the \textit{de facto} estimate at time $\tmax$ can be estimated as a linear combination of \textit{de jure} MAR-based estimates at times $t\le\tmax$.
To find the standard error, we could ignore uncertainty in the $\hat{\alpha}_t$ and use the standard error of the linear combination, or we could do better by applying the delta approximation method \citep{Oehlert1992} as done by \citet{LiuPang16}.

\section{Simulation}

We performed a simulation study to demonstrate equivalence of the RBI methods and the proposed causal model for estimating the treatment effect at the final time point, and to assess the impact of mis-specification of $\boldsymbol{K}_t$ and $\boldsymbol{\beta}_{t}(t)$.

\subsection{Design}


We consider a RCT with one baseline observation and two post-baseline follow-ups during the treatment period (that is, $\tmax=2$).
Potential untreated outcomes $\boldsymbol{Y}(0)$ in both arms were generated from a multivariate normal (MVN) distribution with mean $\boldsymbol{\mu}(0)=(10,12,14)^'$ and variance-covariance matrix $AR(1)$ with standard deviation 3 and lag-one correlation $0.5$.
Potential fully-treated outcomes were derived using the causal models
$Y_{1}(1) - Y_{1}(0) = \delta_{1} + u_1$ and
$Y_{2}(2) - Y_{2}(0) = \delta_{2} + u_1$, where $\delta_1=1$ and $\delta_2=2$ are average treatment effects and $u_1$ may vary between participants to give treatment effect heterogeneity.
In the control arm, we assumed that all participants had complete data.
Observed outcomes in the control arm were set equal to untreated potential outcomes.


In the active arm, we assumed that participants either completed the entire period of treatment and provided complete measurements ($D_a=2$), or completed only the first period of treatment and failed to provide measurements at visit 2 ($D_a=1$). The treatment discontinuation mechanism was $\logit{p(D_a=1|Y_0,Y_{1}(1))}=\tau_0+\tau_{1}Y_{1}(1)$.
Observed outcomes were set equal to the potential fully-treated outcomes.
We used two values of $\tau_{1}$: $0$ for missing completely at random (MCAR) and $1$ for MAR.
$\tau_0$ was chosen to give $p(D_a=1)=0.5$.


We explored the implications of differences between $\boldsymbol{\beta}_{t}(\tmax)$ and $\boldsymbol{\beta}_{t}(0)$.
Since missing data occur only in the $D_a=1$ group, we needed only consider differences for $t=1$, that is, between $\boldsymbol{\beta}_1(2)$ and $\boldsymbol{\beta}_1(0)$.
The distribution above for $\boldsymbol{Y}(0)$ implies $\boldsymbol{\beta}_1(0)=(0, 0.50)'$.
We used two treatment effect models:
homogeneity, $u_1=0$ implying $\boldsymbol{\beta}_1(2)=(0, 0.50)'=\boldsymbol{\beta}_1(0)$;
and heterogeneity, $u_1 \sim N(0,2.5^2)$ implying $\boldsymbol{\beta}_1(2)=(-0.12,0.74)' \ne \boldsymbol{\beta}_1(0)$.
The latter model resulted in slightly more heterogeneity in the active arm than in the control arm.
We considered 250 participants per arm and 1000 repetitions for the simulation study.


This completes the mechanism for simulating the observed data.
However, the \textit{de facto} estimand depends on the unobserved data  including the potential partly-treated outcomes $Y_2(1)$ whose distribution has not yet been specified. We generated $Y_2(1)$ from the causal model $Y_2(1)-Y_2(0)=k \delta_1+u_2$ where $k$ took values 0, 0.5, 0.74 or 1, and where the marginal distribution of $u_2$ was the same as that of $u_1$ with either
(a) $corr(u_1,u_2)=0.5$, so that $\boldsymbol{\beta}_1(1)=\boldsymbol{\beta}_1(0)$,
or
(b) $corr(u_1,u_2)=1$, so that $\boldsymbol{\beta}_1(1)=\boldsymbol{\beta}_1(2)$.
For the treatment effect homogeneity model, options (a) and (b) are equivalent.


For each mechanism for simulating the observed data, we analysed the data in three ways.
Firstly (reported in panel A of Tables \ref{tab3} and \ref{tab4}), we generated the complete data by setting $Y_2=Y_2(D_a)$ under the various values of $k$ and the two choices of $\boldsymbol{\beta}_1(1)$ described above, and analysed the complete data.
Secondly (reported in panel B of Tables \ref{tab3} and \ref{tab4}) we imputed the missing data using the causal model.
Here $\boldsymbol{K}_1$ was a $1 \times 2$ vector whose first element was irrelevant and whose second element was the scalar $k$.
Using $\tilde{k}$  and  $\tilde{\boldsymbol{\beta}}_1(1)$ to denote assumed values of $k$ and $\boldsymbol{\beta}_1(1)$, we took the same range of values for $\tilde{k}$ as proposed for $k$ above, and the same choices of $\tilde{\boldsymbol{\beta}}_1(1)$.
Thirdly (reported in panel C of Tables \ref{tab3} and \ref{tab4}) we imputed the missing data using the reference-based imputation methods CR, CIR and J2R with the two choices of variance-covariance matrix.
In all cases, we estimated the treatment effect from a linear regression of $Y_2$ on randomised arm and baseline $Y_0$. With imputed data, standard errors were computed using Rubin's rules.

\subsection{Results}

Table \ref{tab3} displays the average estimated treatment difference at the final visit for each data generating mechanism.

Comparing panels A and B shows that the causal model imputation methods result in unbiased estimates when the assumed values of $\tilde{\boldsymbol{\beta}}_1(1)$ and $\tilde{k}$ agree with the true values of ${\boldsymbol{\beta}}_1(1)$ and ${k}$.

Comparing panels B and C shows that the RBI estimates with variance-covariance matrix drawn from the control arm are special cases of the causal model estimates with $\boldsymbol{\beta}_1(1)=\boldsymbol{\beta}_1(0)$,
and the RBI estimates with variance-covariance matrix drawn from the active arm are special cases of the causal model estimates with $\boldsymbol{\beta}_1(1)=\boldsymbol{\beta}_1(2)$.
Specifically,
J2R corresponds to $\tilde{k}=0$,
CIR corresponds to $\tilde{k}=1$,
and CR corresponds to $\tilde{k}=$ the second element of $\boldsymbol{\beta}_1(1)$ which is 0.50 or 0.74 depending on the data generating mechanism.

Comparing different choices of $\tilde{\boldsymbol{\beta}}_1(1)$ when the observed data had either no selection effect (i.e.\ under MCAR) or $\boldsymbol{\beta}_1(0)=\boldsymbol{\beta}_1(2)$ (that is, in the first three data generating mechanisms), we see that choice of $\tilde{\boldsymbol{\beta}}_1(1)$ does not affect estimates, as expected from Section \ref{sec:methods:comp}.
Sensitivity to choice of $\tilde{\boldsymbol{\beta}}_1(1)$ was observed in the fourth data generating mechanism (MAR with $\boldsymbol{\beta}_1(0)\ne\boldsymbol{\beta}_1(2)$): mean causal model estimates were reduced by 0.29 by assuming $\boldsymbol{\beta}_1=\boldsymbol{\beta}_1(0)$ instead of $\boldsymbol{\beta}_1=\boldsymbol{\beta}_1(2)$.
Sensitivity to choice of $\tilde{k}$ was the same for all values of $\boldsymbol{\beta}_1$ (see Section \ref{sec:methods:comp}): for example, assuming $\tilde{k}=0$ instead of $\tilde{k}=1$  reduced mean causal model estimates by 0.50 irrespective of the value of $\boldsymbol{\beta}_1$.

\begin{center}
[TABLE \ref{tab3} ABOUT HERE]
\end{center}

Table \ref{tab4} displays the average standard error (SE) (the average of the 1000 SEs) and the empirical SE (the sample standard deviation of the 1000 point estimates) for the treatment difference at the final visit.
Empirical and average SEs for J2R and CIR are similar to those for corresponding causal model estimates. The SEs for CR are slightly larger than that for causal model with $\tilde{k}=$ the second element of $\boldsymbol{\beta}_1$ because $\tilde{k}$ is an assumed value while $\boldsymbol{\beta}_1$ is estimated. With MAR data, the larger average and empirical SEs due to using the variance-covariance matrix from the active arm rather than from the control arm is mainly because of no missing data in the control arm and larger heterogeneity in the active arm than in the control arm. More importantly, as shown in \citet{ian:Carpenter_letter}, the results confirm that both RBI and causal model methods give (1)  smaller empirical SEs than the estimator based on the complete data, and (2) larger average SEs (estimated using RR) than the empirical SEs of the methods and the empirical SEs based on the complete data. We comment on these observations in the discussion.

\begin{center}
[TABLE \ref{tab4} ABOUT HERE]
\end{center}

\section{Examples} \label{sec:examples}

We use two example data sets from randomised, double blind, parallel-group studies comparing active drug with placebo.
The first is from a trial of 172 participants with major depressive disorders, taken from the DIA page of \href{url}{www.missingdata.org.uk}, and used in the DIA working group to demonstrate various missing data related analytical methods. The outcome variable is the 17-item Hamilton Depression Rating Scale, HAMD17.
The second, kindly supplied by Devan Mehrotra, is from a pain trial with a pain score as outcome.

\begin{figure}[b]
\begin{center}
\caption{HAMD17 and pain score data sets: observed mean profile according the time at which treatment discontinued in the drug and placebo arms.}
\label{fig1}
\centering
\includegraphics{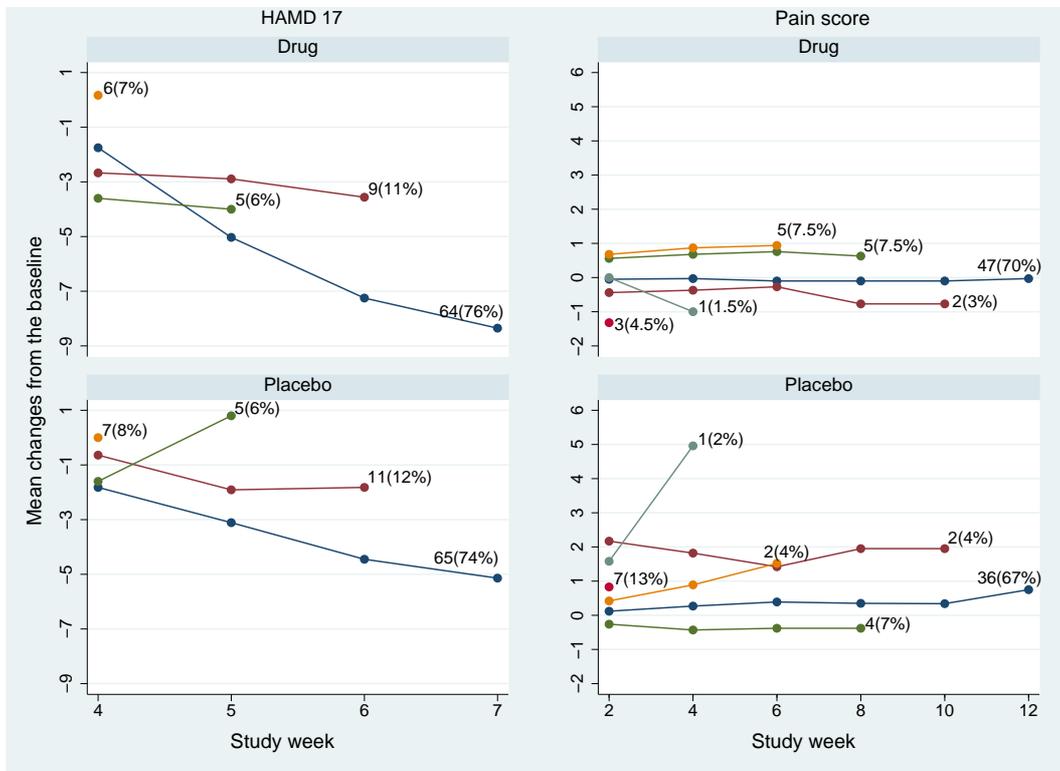}
\begin{minipage}{1.0\textwidth} 
{\footnotesize Note: In the pain score data, four subjects in the drug arm and two subjects in the placebo arm did not complete any post-baseline visit and were excluded from analysis.\par}
\end{minipage}
\end{center}
\end{figure}

In the HAMD17 trial, 76\% (64/84) and 74\% (65/88) of randomised participants completed the final (fourth) visit in the drug and the placebo arms respectively. In the pain score trial, the completion rate at the final (sixth) visit was 70\% (47/67) and 67\% (36/54) in the drug and the placebo arms respectively. In both trials, participants were not followed up after treatment discontinuation. The observed trajectory means and the frequency of dropout patterns in each trial are shown in Figure \ref{fig1}.

\begin{center}
[FIGURE \ref{fig1} ABOUT HERE]
\end{center}

We used the SAS 5 macros for implementing the RBI and the causal models (Section \ref{sssec:sasmacro}).
For the RBI methods, we assumed participants in the drug arm were treated similarly to the placebo arm after discontinuing the drug. To construct the joint distribution of pre- and post-discontinuation drug-arm data under the RBI methods, we first used the variance-covariance matrix from the placebo arm (RBI1 analyses) and then repeated the methods with the variance-covariance matrix from the drug arm (RBI2 analyses).

Table \ref{tab5} shows the estimated treatment effect on HAMD17 and pain score at the final visit from standard MI, MMRM and RBI methods. The standard MI and MMRM methods estimate the \textit{de jure} estimand. These differ slightly for HAMD17 because of a small incompatibility between the imputation and analysis models: the imputation model used all visits to estimate a common the effect of the baseline covariate \texttt{PoolInv}, but the analysis model used only the final visit.
The RBI methods estimate the \textit{de facto} estimand and showed, as expected,  treatment estimates of smaller magnitude than the \textit{de jure} estimand, with J2R giving the smallest magnitude of treatment effect followed by CR. Using the variance-covariance matrix from the drug arm rather than from the placebo arm gave slightly more conservative estimates.

\begin{center}
[TABLE \ref{tab5} ABOUT HERE]
\end{center}

\begin{figure}[b]
\begin{center}
\caption{HAMD17 and pain score data sets: tipping point analysis for the estimated treatment effect at the final visit using causal model (\ref{eq:mte1}) with sensitivity parameter $k_0$.
The horizontal solid and dotted lines represent the treatment estimates and their pointwise 95\% CI, respectively. The vertical solid line corresponds to $k_0$ such that p-value $> 0.05$ in the left hand side of the line (tipping point).}
\label{fig2}
\includegraphics{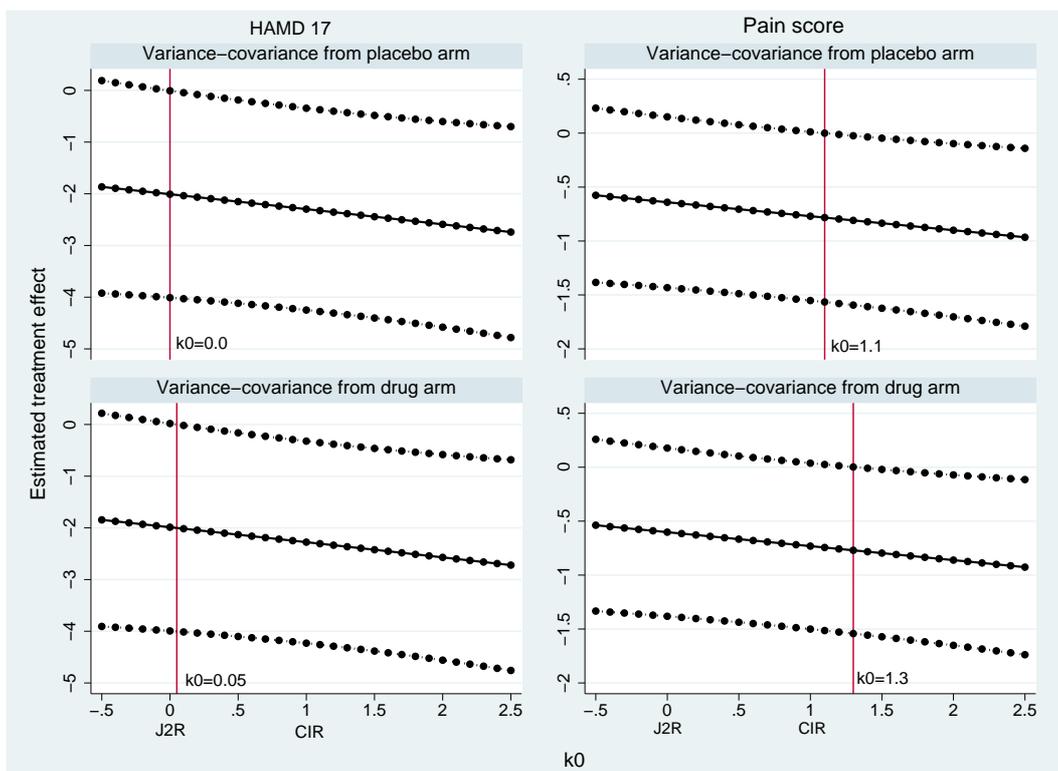}
\end{center}
\end{figure}

We demonstrate tipping point sensitivity analyses using causal models (\ref{eq:mte1}) and (\ref{eq:mte2}).
In  model (\ref{eq:mte1}), a fraction $k_0$ of the treatment effect is maintained at all times after discontinuation.
Figure \ref{fig2} shows the \textit{de facto} estimates and 95\% CI over a range of $k_0$ from -0.5 to 2.5. As shown in the theory and the simulation results, the J2R and CIR estimates correspond to using the causal model with $k_0=0$ (i.e., when participants in the active arm lost their treatment effect after the discontinuation) and $k_0=1$ (i.e., when participants in the active arm maintained their treatment effect after the discontinuation), respectively. Values $k_0<0$ mean that the effect of treatment after discontinuation is harmful, while values $k_0>1$ mean that the effect of treatment after discontinuation is greater than before discontinuation.
The tipping point analysis on HAMD17 shows that statistical significance is lost when $k_0 < 0$ (with variance-covariance from the placebo arm) or $k_0<0.05$ (with variance-covariance from the active arm). In both cases, this suggests that the \textit{de facto} estimate of treatment effect on HAMD17 is non-significant only if any benefit of the active treatment is lost immediately following discontinuation. For the pain score, statistical significance is lost when $k_0<1.1$ or $k_0<1.3$ (depending on whether the variance-covariance matrix is taken from the placebo or the drug arm, respectively). This suggests that, in order for the \textit{de facto} estimate of treatment effect to be statistically significant, there would need to be a delayed benefit such that the treatment effect was greater after discontinuation than before discontinuation.
In both trials, comparing Table \ref{tab5} with Figure \ref{fig2} shows that the MAR analyses give estimates of the \emph{de jure} estimand that are numerically similar to the causal model estimates of the \emph{de facto} estimand when values around $k_0=2$ are assumed.

\begin{center}
[FIGURE \ref{fig2} ABOUT HERE]
\end{center}

\begin{figure}[b]
\begin{center}
\caption{HAMD17 and pain score data sets: tipping point analysis for the estimated treatment effect at the final visit using causal model (\ref{eq:mte2}) with sensitivity parameter $k_1$.
The horizontal solid and dashed lines represent the treatment estimates and their pointwise 95\% CI, respectively. The tipping point is not attained in the range $0 \le k_1 \le 1$.}
\label{fig3}
\includegraphics{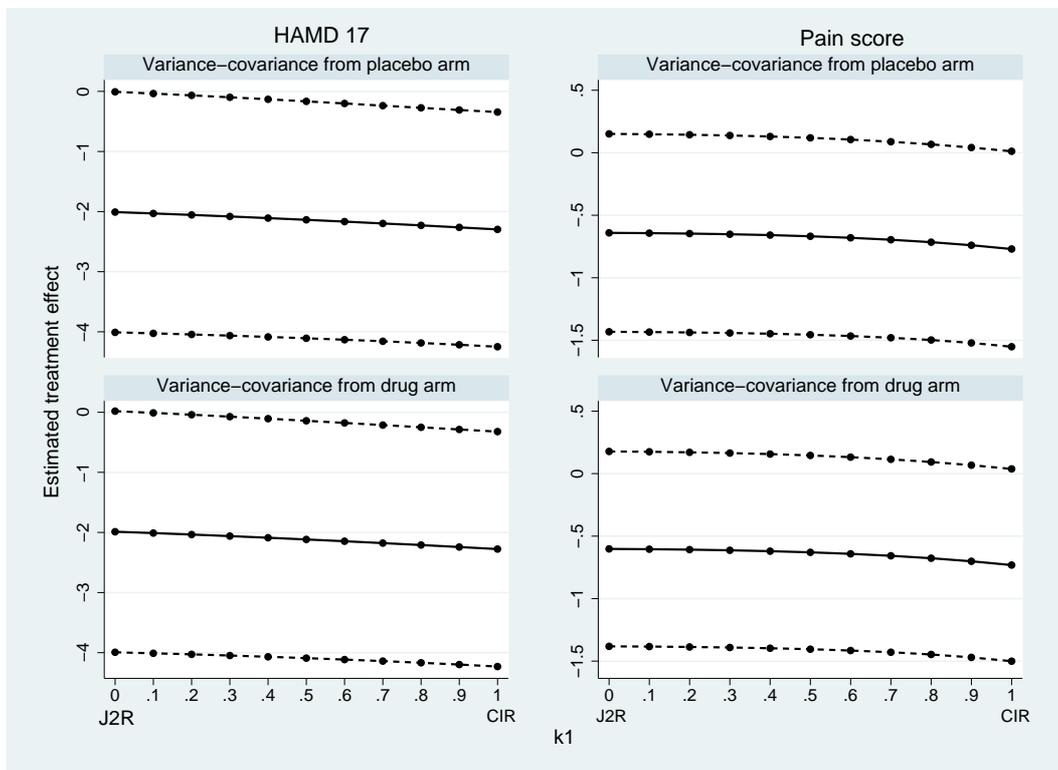}
\end{center}
\end{figure}

In  model (\ref{eq:mte2}), the treatment effect decays exponentially after discontinuation. Here, $k_1 = 0$ for J2R and $1$ for CIR.
Figure \ref{fig3} shows the \textit{de facto} estimates of treatment effect at the final visit and its 95\% CI from the causal model over a range of $k_1$. This model does not accommodate the effect of treatment after discontinuation being either harmful or greater than before discontinuation, and because of the more limited range of $k_1$, the tipping point is not reached: all results are statistically significant for HAMD17 and not significant for the pain score.

\begin{center}
[FIGURE \ref{fig3} ABOUT HERE]
\end{center}

\section{Discussion}
We have considered longitudinal RCTs with quantitative outcomes in which participants who discontinue an active treatment are not followed up thereafter, but are assumed to receive a standard treatment which is often the control treatment. We have focused on estimating the effect of assignment to treatment in the actual treatment circumstances of the trial (\textit{de facto} estimand) rather than the treatment effect if all participants had tolerated or adhered to trial protocol (\textit{de jure} estimand). We have investigated the underlying assumptions of the RBI methods and proposed a generalised causal modelling approach to account for treatment discontinuation in the estimation of the \textit{de facto} estimand. The proposed causal model makes an explicit assumption about the causal effect of a given treatment history, and provides flexibility to perform sensitivity analyses to the causal assumption.

The proposed causal model specifies how much of the treatment effect is maintained after treatment discontinuation, which we represent by the parameter matrix $\boldsymbol{K}_t$.
We illustrated this with two examples of $\boldsymbol{K}_t$: (\ref{eq:mte1}) with the maintained treatment effect independent of time since discontinuation, and (\ref{eq:mte2}) with the maintained treatment effect decaying exponentially with visits since discontinuation. A simple extension would allow $\boldsymbol{K}_t$ to depend on the reason for treatment discontinuation. Ideally sponsors should justify the choice of $\boldsymbol{K}_t$ in the trial protocol based on the nature of the trial and the treatments.

The choice of $\boldsymbol{\beta}_t(t)$ (reflecting within-subject dependence of post-discontinuation outcomes on pre-discontinuation outcomes) should similarly be pre-specified; it is hard to recommend a single choice and perhaps both $\boldsymbol{\beta}_t(t)=\boldsymbol{\beta}_{t}(0)$ and $\boldsymbol{\beta}_t(t)=\boldsymbol{\beta}_{t}(\tmax)$ should be implemented.
If the analyst is willing to assume equal variance-covariance matrices across trial arms then the situation is simpler and $\boldsymbol{\beta}_t(t)=\boldsymbol{\beta}_{t}(0)=\boldsymbol{\beta}_{t}(\tmax)$ is the obvious choice.

Our model has been presented for the case of active arm treatment discontinuation, where subjects who discontinue do not then receive rescue medication (in the case of placebo controlled trials), or only receive standard of care (in the case where the control arm is standard of care).  It can be extended to handle the control arm starting active treatment: an assumption like \ref{ass:del} still holds, but the $\boldsymbol{K}_t$ matrix
must be replaced by assumptions about how the treatment effect builds up over time. 
We are currently working on extending the model to trials comparing two active treatments. An unresolved problem is how to handle initiation of rescue medications. 

We have focussed on varying assumption \ref{ass:del}, but we should also assess a number of other assumptions. The MAR assumptions  \ref{ass:refMAR} and \ref{ass:actMAR1}, and the related assumption \ref{ass:actMAR2}, could be made more plausible if the model could be extended to include further time-dependent covariates.
Alternatively one could explore sensitivity to these assumptions by methods like those of \citet{Ratitch++13}. It is less clear how to assess departures from the linearity assumption \ref{ass:linear}.

We have focussed on quantitative outcomes. Extension to other outcomes would be useful, and we are currently working on a causal version of the method of \citet{Keene2014} for recurrent events.

We proposed approximations to the RR variance in options 2 and 3 of Section \ref{ssec:impl}. From the simulation and empirical results, we see that the SEs are not much affected by the value of $\boldsymbol{K}_t$. This means that our approximations are adequate if we want the RR variance but not if we want the repeated sampling variance.
The repeated sampling variance of the estimated treatment effect tends to be smaller than the RR estimate of variance for a given $\boldsymbol{K}_t$.
The repeated sampling variance can be approximated in practice using the delta method \citep{Oehlert1992,LiuPang16}.
\citet{Carpenter2014} argue that the repeated sampling variance is not appropriate, since it is typically smaller than the complete-data variance (to an extent which strongly depends on the value of $\boldsymbol{K}_t$).
They also argue that the RR estimate of variance of the treatment effect is larger than the the complete-data variance, because of the information lost due to the missing data, and this makes it an appropriate variance \citep{Carpenter2014,Cro++15}.
We point out that the type I error rate is correct for the repeated sampling variance and too small for the RR variance, meaning that the RR variance carries a loss of power: therefore the repeated-sampling variance
may be appropriate for a primary analysis.


In summary, whilst multiple imputation is an attractive and powerful method for handling missing data in both experimental and observational studies, it is not always clear what estimand is being targeted or what assumptions are being made about how outcomes for subjects who discontinue randomised treatment relate to those who remain on study. The recent estimands debate \citep{ICHE9} has led to a growing recognition that more complex estimation approaches that do not rely on randomisation are needed to handle post-randomisation events that lead to missing data, and there are calls for causal inference methods to become more widely adopted (e.g.\ \citet{Akacha2016,LittleKang15}). We join this call to encourage greater understanding and application of ideas from the causal inference literature to help support the definition and estimation of estimands of interest in a randomised clinical trial. We hope that this paper illustrates how a causal inference framework can provide clarity and rigour in stating estimands, stating assumptions, and performing estimation.

\section*{Acknowledgements}
    IRW and RJ were supported by the Medical Research Council (Unit Programme number U105260558).
    RJ was also supported by GSK.
    Nicky Best is an employee and shareholder of GSK.
    We thank James Roger for very helpful comments on an earlier draft and for allowing us to modify his software,
    and Devan Mehrotra for providing the pain score dataset.


\appendix
\renewcommand{\thesection}{Appendix \Alph{section}} 
\newpage
\section{Proof of equation (\ref{eq:dfest})} \label{app:defacto}

We first write the \textit{de facto} estimand at time $\tmax$ as a weighted average over treatment discontinuation times:
\begin{equation}\label{eq:wtave}
E[{Y}_\tmax(D_a)-{Y}_\tmax(0)]
= \sum_{t=0}^{\tmax} \alpha_t \left\{ \mu_\tmax(t | D_a=t) - {\mu}_\tmax(0) \right\}
\end{equation}
where we define $\boldsymbol{\mu}(s|D_a=s) = E[\boldsymbol{Y}(s) | D_a=s]$ etc.

We next write the terms on the right hand side of (\ref{eq:wtave}) for $t<\tmax$ using the result in Table \ref{tab:imputed}:
\begin{eqnarray}
\mu_\tmax(t | D_a=t) - \mu_\tmax(0)
&=& \boldsymbol{e}_t' \boldsymbol{\beta}_{t}(t)
\left\{\boldsymbol{\mu}_{\pret}(t|D_a=t) - \boldsymbol{\mu}_{\pret}(t)\right\}\nonumber\\
&& +\ \boldsymbol{e}_t' \boldsymbol{K}_t \left\{\boldsymbol{\mu}_{\pret}(t)-\boldsymbol{\mu}_{\pret}(0)\right\} \label{eq:app0}
\end{eqnarray}
where $\boldsymbol{e}_t$ is used to extract the $\tmax$th element.

Selection effects cancel out over treatment discontinuation times $t$. To use this, note that \ref{ass:linear} implies
\begin{eqnarray*}
\boldsymbol{\beta}_{t}(\tmax) \left\{\boldsymbol{\mu}_{\pret}(t|D_a=t) - \boldsymbol{\mu}_{\pret}(t)\right\}
&=& \boldsymbol{\mu}_{\postt}(\tmax|D_a=t) - \boldsymbol{\mu}_{\postt}(\tmax)
\end{eqnarray*}
for each $t < \tmax$, and the sum of the right hand side over $t$ (weighted by $\alpha_t$) is zero. We sum the left hand side (weighted by $\alpha_t$) and take the last element, giving
\begin{eqnarray*}
&&\hspace{-3 em}\sum_{t<\tmax} \alpha_t \boldsymbol{e}_t' \boldsymbol{\beta}_{t}(\tmax)
\left\{\boldsymbol{\mu}_{\pret}(t|D_a=t) - \boldsymbol{\mu}_{\pret}(t)\right\}\\
&&+\ \alpha_\tmax \left\{ {\mu}_\tmax(\tmax|D_a=\tmax)- {\mu}_\tmax(\tmax) \right\} =0
\end{eqnarray*}
and hence
\begin{eqnarray}\label{eq:lemma2}
&&\hspace{-3 em}\alpha_\tmax {\mu}_\tmax(\tmax|D_a=\tmax)
=
\alpha_\tmax {\mu}_\tmax(\tmax) \nonumber\\
&&\hspace{3 em}-
\sum_{t<\tmax} \alpha_t \boldsymbol{e}_t' \boldsymbol{\beta}_{t}(\tmax)
\left\{\boldsymbol{\mu}_{\pret}(t|D_a=t) - \boldsymbol{\mu}_{\pret}(t)\right\}.
\end{eqnarray}

Finally, substituting equations (\ref{eq:lemma2}) and (\ref{eq:app0}) into (\ref{eq:wtave}) gives
\begin{eqnarray*}
E[{Y}_\tmax(D)-{Y}_\tmax(0)]
&=&\alpha_\tmax \left\{ {\mu}_\tmax(\tmax)-{\mu}_\tmax(0) \right\} \\
&&+ \sum_{t<\tmax} \alpha_t \boldsymbol{e}_t' (\boldsymbol{\beta}_{t}(t)-\boldsymbol{\beta}_{t}(\tmax))
\left\{\boldsymbol{\mu}_{\pret}(t|D_a=t) - \boldsymbol{\mu}_{\pret}(t)\right\} \\
&&+ \sum_{t<\tmax} \alpha_t \boldsymbol{e}_t' \boldsymbol{K}_t \left\{\boldsymbol{\mu}_{\pret}(t)-\boldsymbol{\mu}_{\pret}(0)\right\}
\end{eqnarray*}
which is equivalent to equation (\ref{eq:dfest}) in the main text.

\section{Causal model implementation using the SAS 5 macros} \label{app:code}


The modified macro file `Part2A\_38\_causal.sas' and the HAMD17 data are available on the DIA section of the website \href{url}{www.missingdata.org.uk}.
The HAMD17 data are stored as a SAS data set named \texttt{Chapter15\_example} which we assume has been renamed \texttt{HAMD}.

The modified \texttt{Part2A} macro allows the user to impute under the causal model by specifying option \texttt{Method=Causal}. The value of $k$ in equation (\ref{eq:Kt}) is specified by \texttt{Causalk=} or defaults to 1 giving the CIR estimate. $k$ can be specified as a scalar (a constant $k$ for all participants; e.g.\ \texttt{Causalk=0.5}) or as a variable that specifies $k$ for each participant (e.g.\ \texttt{Causalk=varname}). If $k$ is specified using a variable, that variable must be identified in macro Part1A using the option \texttt{id=varname}.
If \texttt{Method=Causal} is not specified then RBI methods are implemented and \texttt{Causalk=} is ignored.

Causal model (\ref{eq:mte1}) with $k_0=0.5$ can be fitted to the HAMD17 data by:
\begin{verbatim}
%Part1A(Jobname=Example, Data=HAMD, Subject=Patient, Response=Change,
        Time=Visit, Treat=Therapy, Catcov=PoolInv, Covbytime=Basval,
        Covgroup=Therapy);
%Part1B(Jobname=Example, Ndraws=100, Thin=100, Seed=12345);
%Part2A(Jobname=Example_causal, Inname=Example, Method=Causal,
        Causalk=.5, Ref=Placebo, VCMethod=Ref);
\end{verbatim}

The option \texttt{Covgroup=Therapy} in the \texttt{\%Part1A} call specifies that the variance-covariance matrix is estimated separately by arm, but \texttt{Catcov=PoolInv} and \texttt{Covbytime=Basval} specify that the coefficients of the baseline covariates are assumed equal across arms.

Causal model (\ref{eq:mte2}) with $k_1=0.5$ can be fitted to the HAMD17 data after creating a new variable \texttt{k1power} to hold the individual's value of $k_1^{v_s-v_t}$ where $s=\tmax$ and $t=D_a$. In these data \texttt{Visit} is coded 0, 4, 5, 6, 7, so we set $v_s =7$ and compute $v_t$ as the last value of \texttt{Visit}:
\begin{verbatim}
proc sql;
create table HAMD2 as
   select *, 0.5**(7-max(Visit)) as k1power
   from HAMD
   group by Patient;
quit;
\end{verbatim}
Although \texttt{k1power} is computed for all individuals, the next steps ignore values computed in the $Z=c$ arm:
\begin{verbatim}
%Part1A(Jobname=Example, Data=HAMD2, Subject=Patient, Response=Change,
        Time=Visit, Treat=Therapy, Catcov=PoolInv, Covbytime=Basval,
        Covgroup=Therapy, Id=k1power);
%Part1B(Jobname=Example, Ndraws=100, Thin=100, Seed=12345);
%Part2A(Jobname=Example_causal, Inname=Example, Method=Causal,
        Causalk=k1power, Ref=Placebo, VCMethod=Ref);
\end{verbatim}

\texttt{\%Part1A} and \texttt{\%Part1B} need not be repeated for different values of $k_0$ or $k_1$.
The imputed data sets in either case are analysed using
\begin{verbatim}
%Part2B(Jobname=Example_causal, Seed=54321);
%Part3(Jobname=Example_causal, Anref=Placebo,
        Label=Causal with sigma from reference);
\end{verbatim}


\bibliographystyle{\home/latex/bst/jrss} 



\clearpage
\renewcommand{\baselinestretch}{1}

\begin{table}[tp]
\begin{center}
\caption{Notation.}
\label{tab:notation}
\begin{tabular}{p{5cm} p{8cm}}
\hline
Notation & Explanation \\ [0.5ex]
\hline
$Z$ & Randomised treatment arm: $Z=a$ for the active treatment and $Z=c$ for the control treatment. \\
$D_a$ or $D_c$ & Participant's last observation time prior to stopping treatment in arm $a$ or $c$. \\
$Y_t$ & Response at time $t$, where $t =0,...,\tmax$. \\
$\boldsymbol{Y}=(\boldsymbol{Y}_{\pret},\boldsymbol{Y}_{\postt})$  &
The vector of all outcomes from $t=0$ to $t=\tmax$, with subvectors
for outcomes until time $t$, and
for outcomes from time $t+1$. \\
$Y_t(s)$ & Potential outcome at time $t$ if  $s$ periods of the active treatment were received, where $0 \le s \le \tmax$. \\
$\boldsymbol{Y}(s)=(\boldsymbol{Y}_{\pret}(s),\boldsymbol{Y}_{\postt}(s))$  & The vector of all potential outcomes if $s$ periods of the active treatment were received, with subvectors
for outcomes until time $t$, and
for outcomes from time $t+1$. \\
$\boldsymbol{\mu}(s)$ & Vector of overall means of counterfactuals if participants would receive only $s$ periods of the active treatment. \\
$\boldsymbol{\Sigma}(s)$ & Variance-covariance matrix  of $\varb{\boldsymbol{Y}(s)}$ for $s=0,\tmax$. \\
$\boldsymbol{\beta}_t(s)$ & Regression coefficient of $\boldsymbol{Y}_{\postt}(s)$ on $\boldsymbol{Y}_{\pret}(s)$ for $s=0,\tmax$. \\
$\boldsymbol{\Sigma}_{\postt\postt|\pret}(s) $ & Residual variance of $\boldsymbol{Y}_{\postt}(s)$ given $\boldsymbol{Y}_{\pret}(s)$ for $s=0,\tmax$. \\[1ex]
\hline
\end{tabular}
\end{center}
\end{table}
\clearpage


\begin{table}[tp]
\begin{center}
\caption{Mean of the imputation distribution of $\boldsymbol{Y}_{\postt}(t)$ for $t<\tmax$ given randomisation $Z=a$, past $\boldsymbol{Y}_{\pret}$ and treatment discontinuation time $D_a=t$, under various reference-based imputation methods \citep{Carpenter++13} and the proposed causal model.}
\label{tab:imputed}
\begin{tabular}{llll}\hline
Method&	\multicolumn{3}{l}{Mean of the imputation distribution}\\\hline
\multicolumn{4}{l}{\emph{Reference-based imputation methods}}\\
LMCF & $\boldsymbol{\beta}_{t}(\tmax)
\left\{\boldsymbol{Y}_{\pret} - \boldsymbol{\mu}_{\pret}(t)\right\} $
&$+ \boldsymbol{C}_t\boldsymbol{\mu}_{\pret}(t)$	\\
J2R\textsuperscript{*}
& $\boldsymbol{\beta}_{t}(0)
\left\{\boldsymbol{Y}_{\pret} - \boldsymbol{\mu}_{\pret}(t)\right\}$
&
&$+ \boldsymbol{\mu}_{\postt}(0)$\\
CIR\textsuperscript{*}	 & $\boldsymbol{\beta}_{t}(0)
\left\{\boldsymbol{Y}_{\pret} - \boldsymbol{\mu}_{\pret}(t)\right\} $
&$+ \boldsymbol{C}_t\left\{\boldsymbol{\mu}_{\pret}(t)-\boldsymbol{\mu}_{\pret}(0)\right\}$
&$+ \boldsymbol{\mu}_{\postt}(0)$ \\
CR\textsuperscript{*}	 & $\boldsymbol{\beta}_{t}(0)
\left\{\boldsymbol{Y}_{\pret} - \boldsymbol{\mu}_{\pret}(t)\right\} $
&$+ \boldsymbol{\beta}_{t}(0)
\left\{\boldsymbol{\mu}_{\pret}(t) - \boldsymbol{\mu}_{\pret}(0)\right\}$
&$+ \boldsymbol{\mu}_{\postt}(0)$	\\\hline
\emph{Causal model} & $\boldsymbol{\beta}_{t}(t) \left\{\boldsymbol{Y}_{\pret} - \boldsymbol{\mu}_{\pret}(t)\right\} $
&$+ \boldsymbol{K}_t \left\{\boldsymbol{\mu}_{\pret}(t)-\boldsymbol{\mu}_{\pret}(0)\right\}$
&$+ \boldsymbol{\mu}_{\postt}(0)$ \\ \hline
\multicolumn{4}{l}{\footnotesize{*with control arm as the reference and variance-covariance from the control arm}}
\end{tabular}
\end{center}
\end{table}
\clearpage

\begin{table}[t]
\begin{center}
\vspace{-5ex}
\caption{Simulation study: estimates of treatment effect at the final visit using complete data, causal model imputation and RBI imputation. $\boldsymbol{\beta}_1(0)=(0,0.5)'$ in all cases.
$\boldsymbol{\beta}_1(2) \ne \boldsymbol{\beta}_1(0)$ means $\boldsymbol{\beta}_1(2)=(-0.12,0.74)'$.}
\label{tab3}
\begin{tabular}{lcccc}
\hline \\[-1.9ex]
\multirow{4}{*}{} & \multicolumn{4}{c}{Data generating mechanisms for observed data}           \\ \cline{2-5} \\[-1.9ex]
   & \multicolumn{2}{c}{MCAR}    & \multicolumn{2}{c}{MAR}     \\ \cline{2-5} \\[-1.9ex]
	   & $\boldsymbol{\beta}_1(2)=\boldsymbol{\beta}_1(0)$ & $\boldsymbol{\beta}_1(2)\ne \boldsymbol{\beta}_1(0)$ & $\boldsymbol{\beta}_1(2)=\boldsymbol{\beta}_1(0)$ & $\boldsymbol{\beta}_1(2)\ne \boldsymbol{\beta}_1(0)$ \\ \hline \\[-1.5ex]
\multicolumn{5}{l}{\textbf{A}. Complete data generated with:}   \\
$\boldsymbol{\beta}_1(1)=\boldsymbol{\beta}_1(0)$  &   &     &   &     \\
$\quad k=0.00$  & 0.99   & 0.99     & 1.00   & 0.70     \\
$\quad k=0.50$  & 1.24   & 1.24     & 1.25   & 0.95     \\
$\quad k=0.74$  & 1.36   & 1.36     & 1.37   & 1.07     \\
$\quad k=1.00$  & 1.49   & 1.49     & 1.50   & 1.20     \\
$\boldsymbol{\beta}_1(1)=\boldsymbol{\beta}_1(2)$ & & && \\
$\quad k=0.00$  & 0.99   & 1.00     & 1.00   & 1.00     \\
$\quad k=0.50$  & 1.24   & 1.25     & 1.25   & 1.25     \\
$\quad k=0.74$  & 1.36   & 1.37     & 1.37   & 1.37     \\
$\quad k=1.00$  & 1.49   & 1.50     & 1.50   & 1.50     \\
\hline \\[-1.5ex]
\multicolumn{5}{l}{\textbf{B}. Causal model imputation with assumed $\tilde{\boldsymbol{\beta}}_1(1)$ and $\tilde{k}$}           \\
$\tilde{\boldsymbol{\beta}}_1(1)=\boldsymbol{\beta}_1(0)$     &   &     &   &     \\
$\quad\tilde{k}=0.00$     & 1.00   & 1.00     & 1.00   & 0.71     \\
$\quad\tilde{k}=0.50$     & 1.24   & 1.25     & 1.25   & 0.96     \\
$\quad\tilde{k}=0.74$     & 1.36   & 1.37     & 1.37   & 1.08     \\
$\quad\tilde{k}=1.00$     & 1.49   & 1.50     & 1.50   & 1.21     \\
$\tilde{\boldsymbol{\beta}}_1(1)=\boldsymbol{\beta}_1(2)$     &   &     &   &     \\
$\quad\tilde{k}=0.00$   & 1.00   & 1.00     & 1.00   & 1.00     \\
$\quad\tilde{k}=0.50$    & 1.24   & 1.25     & 1.25   & 1.25     \\
$\quad\tilde{k}=0.74$    & 1.36   & 1.37     & 1.37   & 1.37     \\
$\quad\tilde{k}=1.00$    & 1.49   & 1.50     & 1.50   & 1.50     \\
\hline \\[-1.5ex]
\multicolumn{5}{l}{\textbf{C}. RBI imputation with assumed variance-covariance and method}        \\
\multicolumn{3}{l}{Variance-covariance matrix from control arm}  &   &     \\
\quad J2R     & 1.00   & 1.00     & 1.00   & 0.71     \\
\quad CR      & 1.24   & 1.25     & 1.25   & 0.96     \\
\quad CIR     & 1.49   & 1.50     & 1.50   & 1.21     \\
\multicolumn{4}{l}{Variance-covariance matrix from active arm}       &     \\
\quad J2R     & 1.00   & 1.00     & 1.00   & 1.00     \\
\quad CR      & 1.24   & 1.37     & 1.25   & 1.38     \\
\quad CIR     & 1.49   & 1.50     & 1.50   & 1.50     \\
\hline
\multicolumn{5}{l}{\footnotesize{Note: Maximum Monte Carlo standard error $< 0.01$}}
\end{tabular}
\end{center}
\end{table}
\clearpage

\begin{table}[t]
\begin{center}
\vspace{-5ex}
\caption{Simulation study: average standard error (empirical standard error) for the treatment difference at the final visit using complete data, causal model imputation and RBI imputation.
$\boldsymbol{\beta}_1(0)$ and
$\boldsymbol{\beta}_1(2)$ as in Table \ref{tab3}.
\vspace{2 ex}}
\label{tab4}
\begin{tabular}{lcccc}
\hline \\[-1.9ex]
\multirow{4}{*}{} & \multicolumn{4}{c}{Data generating mechanisms for observed data}           \\ \cline{2-5} \\[-1.9ex]
   & \multicolumn{2}{c}{MCAR}    & \multicolumn{2}{c}{MAR}     \\ \cline{2-5} \\[-1.9ex]
& $\boldsymbol{\beta}_1(2)=\boldsymbol{\beta}_1(0)$
& $\boldsymbol{\beta}_1(2)\ne \boldsymbol{\beta}_1(0)$
& $\boldsymbol{\beta}_1(2)=\boldsymbol{\beta}_1(0)$
& $\boldsymbol{\beta}_1(2)\ne \boldsymbol{\beta}_1(0)$ \\ \hline \\[-1.5ex]
\multicolumn{5}{l}{\textbf{A}. Complete data generated with:}   \\
$\boldsymbol{\beta}_1(1)=\boldsymbol{\beta}_1(0)$  &   &     &   &     \\
$\quad k=0.00$  & 0.276 (0.273)	&	0.318 (0.311)	&	0.260 (0.255)	&	0.295 (0.288)     \\
$\quad k=0.50$  & 0.272 (0.269)	&	0.315 (0.308)	&	0.261 (0.256)	&	0.298 (0.291)     \\
$\quad k=0.74$  & 0.302 (0.206)	&	0.342 (0.247)	&	0.311 (0.219)	&	0.351 (0.258)     \\
$\quad k=1.00$  & 0.270 (0.266)	&	0.313 (0.306)	&	0.262 (0.257)	&	0.301 (0.295)    \\
$\boldsymbol{\beta}_1(1)=\boldsymbol{\beta}_1(2)$ & & && \\
$\quad k=0.00$  & 0.276 (0.273) &	0.318 (0.316)	&	0.260 (0.255)	&	0.292 (0.288)    \\
$\quad k=0.50$  & 0.272 (0.269)	&	0.315 (0.312)	&	0.261 (0.256)	&	0.296 (0.291)     \\
$\quad k=0.74$  & 0.271 (0.268)	&	0.314 (0.311)	&	0.262 (0.256)	&	0.298 (0.294)     \\
$\quad k=1.00$  & 0.270 (0.266)	&	0.313 (0.310)	&	0.262 (0.257)	&	0.300 (0.296)     \\
\hline \\[-1.5ex]
\multicolumn{5}{l}{\textbf{B}. Causal model imputation with assumed $\tilde{\boldsymbol{\beta}}_1(1)$ and $\tilde{k}$}           \\
$\tilde{\boldsymbol{\beta}}_1(1)=\boldsymbol{\beta}_1(0)$     &   &     &   &     \\
$\quad\tilde{k}=0.00$     & 0.310 (0.168)	&	0.337 (0.189)	&	0.305 (0.171)	&	0.328 (0.181)     \\
$\quad\tilde{k}=0.50$     & 0.301 (0.190)	&	0.327 (0.226)	&	0.299 (0.189)	&	0.322 (0.214)     \\
$\quad\tilde{k}=0.74$     & 0.302 (0.206)	&	0.327 (0.249)	&	0.301 (0.206)	&	0.325 (0.237)     \\
$\quad\tilde{k}=1.00$     & 0.305 (0.226)	&	0.332 (0.277)	&	0.306 (0.227)	&	0.332 (0.267)     \\
$\tilde{\boldsymbol{\beta}}_1(1)=\boldsymbol{\beta}_1(2)$     &   &     &   &     \\
$\quad \tilde{k}=0.00$   & 0.310 (0.168)	&	0.359 (0.187)	&	0.315 (0.188)	&	0.359 (0.206)     \\
$\quad\tilde{k}=0.50$    & 0.301 (0.190)	&	0.344 (0.224)	&	0.310 (0.204)	&	0.350 (0.236)     \\
$\quad\tilde{k}=0.74$    & 0.302 (0.206)	&	0.342 (0.247)	&	0.311 (0.219)	&	0.351 (0.258)     \\
$\quad\tilde{k}=1.00$     & 0.305 (0.226)	&	0.345 (0.275)	&	0.316 (0.239)	&	0.356 (0.285)     \\
\hline \\[-1.5ex]
\multicolumn{5}{l}{\textbf{C}. RBI imputation with assumed variance-covariance and method}        \\
\multicolumn{4}{l}{Variance-covariance matrix from control arm}  &       \\
\quad J2R     & 0.310 (0.168)	&	0.337 (0.189)	&	0.305 (0.171)	&	0.328 (0.181)     \\
\quad CR      & 0.303 (0.192)	&	0.328 (0.229)	&	0.306 (0.200)	&	0.331 (0.226)     \\
\quad CIR     & 0.305 (0.226)	&	0.332 (0.277)	&	0.306 (0.227)	&	0.332 (0.267)    \\
\multicolumn{4}{l}{Variance-covariance matrix from active arm}       &     \\
\quad J2R     & 0.310 (0.168)	&	0.359 (0.187)	&	0.315 (0.188)	&	0.359 (0.206)    \\
\quad CR      & 0.305 (0.192)	&	0.344 (0.249)	&	0.332 (0.236)	&	0.370 (0.283)     \\
\quad CIR     & 0.305 (0.226)	&	0.345 (0.275)	&	0.316 (0.239)	&	0.356 (0.285)     \\
\hline
\multicolumn{5}{l}{\footnotesize{Note: Maximum Monte Carlo standard error $< 0.0005$}}
\end{tabular}
\end{center}
\end{table}
\clearpage

\begin{table}[t]
\begin{center}
\caption{HAMD17 and pain score data: estimated treatment effect at the final visit using standard multiple imputation with 100 imputations, mixed model for repeated measures (MMRM) and RBI methods.
}
\label{tab5}
\begin{tabular}{lccccccc}
\hline \\[-1.9ex]
\emph{Estimand} \&  & \multicolumn{3}{c}{HAMD17} && \multicolumn{3}{c}{Pain score}  \\ \cline{2-4}\cline{6-8} \\[-1.9ex]
Method             & Estimate$^1$  & Std. error & p-value && Estimate$^2$ & Std. error & p-value \\ \hline \\[-1.5ex]
\emph{De jure}  \\
\quad Standard MI  & -2.62    & 0.99      & 0.01   && -0.88   & 0.39      & 0.03   \\
\quad MMRM         & -2.58    & 1.03      & 0.01   && -0.88   & 0.39      & 0.03   \\ \\[-1.5ex]
\emph{De facto}\\
\multicolumn{8}{l}{RBI1: variance-covariance from the placebo arm} \\
\quad J2R          & -2.01    & 1.01      & 0.05   && -0.64   & 0.40      & 0.11   \\
\quad CR           & -2.22    & 0.99      & 0.03   && -0.75   & 0.39      & 0.06   \\
\quad CIR          & -2.30    & 0.99      & 0.02   && -0.77   & 0.39      & 0.05   \\ \\[-1.5ex]
\multicolumn{8}{l}{RBI2: variance-covariance from the drug arm}    \\
\quad J2R          & -1.99    & 1.01      & 0.05   && -0.60   & 0.39      & 0.13   \\
\quad CR           & -2.20    & 0.99      & 0.03   && -0.71   & 0.39      & 0.07    \\
\quad CIR          & -2.28    & 0.99      & 0.02   && -0.73   & 0.39      & 0.06 \\ \hline
\hline
\end{tabular}
\end{center}
$^1$ Monte Carlo standard error for MI methods is $\le 0.04$.\\
$^2$ Monte Carlo standard error for MI methods is $\le 0.02$.
\end{table}
\clearpage

\clearpage

\clearpage

\clearpage

\end{document}